\documentclass[12pt]{iopart}

\usepackage{graphicx}

\newtheorem{proposition}{Proposition}
\newtheorem{corollary}{Corollary}
\newtheorem{lemma}{Lemma}
\newtheorem{theorem}{Theorem}
\newtheorem{definition}{Definition}
\newtheorem{remark}{Remark}

\begin{document}

\title[Information Geometry and Nonlinear Diffusion Equations]{Information Geometry of $q$-Gaussian Densities 
and Behaviors of Solutions to Related Diffusion Equations \footnote{Preliminary forms of several results in this paper will appear in \cite{Ohara09} 
without proofs.}}

\author{Atsumi Ohara$^1$ and Tatsuaki Wada$^2$}
\address{$^1$ Department of Systems Science, 
Osaka University, Toyonaka, Osaka 560-8531, Japan}
\ead{ohara@sys.es.osaka-u.ac.jp}
\author{}
\address{$^2$ Department of Electrical and Electronic Engineering, Ibaraki University, Hitachi, Ibaraki 316-8511, Japan}
\ead{wada@mx.ibaraki.ac.jp}

\begin{abstract}
This paper presents new geometric aspects of the behaviors 
of solutions to the porous medium equation (PME) and its associated equation.
First we discuss thermostatistical structure with information geometry on 
a manifold of generalized exponential densities.
A dualistic relation between the two existing formalisms 
is elucidated.
Next by equipping the manifold of {\em $q$-Gaussian 
densities} with such a structure, we derive several physically and 
geometrically interesting properties of the solutions.
The manifold is proved invariant and attracting 
for the evolving solutions, which play crutial roles in our analysis.
We demonstrate that the moment-conserving projection 
of a solution coincides with a geodesic curve on the manifold.
Further,  
the evolutional velocities of the second moments and 
the convergence rate to the manifold are evaluated in terms of the 
{\em Bregman divergence}.
Finally we show that the self-similar solution is geometrically special 
in the sense that it is simultaneously geodesic with respect to the 
mutually dual two affine connections.

\end{abstract}

\submitto{Journal of Physics A: Mathematical and Theoretical}
\maketitle

\section{Introduction}
Let $u(x,t)$ and $p(x,\tau)$ 
on ${\bf R}^n \times {\bf R}_+$ be, respectively, the solutions of
the following nonlinear diffusion equation, which is called the
{\em porous medium equation} (PME):
\begin{equation}
 \frac{\partial u}{\partial t}=\Delta u^m, \quad 1<m \in {\bf R}
\label{PME}
\end{equation}
with nonnegative initial data $0 \le u(x,0)=u_0(x) \in L^1(\mathbf{R}^n)$, 
and the associated nonlinear Fokker-Planck equation (NFPE):
\begin{eqnarray}
   \frac{\partial p}{\partial \tau} = \nabla \cdot 
	\left( \beta x p + D \nabla p^m \right), 
		\quad 0<\beta \in {\bf R} 
\label{DNFPE}
\end{eqnarray}
with nonnegative initial data $0 \le p(x,0)=p_0(x) \in L^1(\mathbf{R}^n)$.
Here, $D$ is a real symmetric positive definite matrix, which represents 
the diffusion coefficients.
As is widely known \cite{Vaz,Toscani} and shown later, 
one solution is obtained from 
a simple transformation from the other, and vice versa.


The PME and NFPE with $m>1$ represent the 
{\em slow diffusion} phenomena, 
which naturally arise in many physical problems including 
percolation of fluid through porous media, intensive thermal waves 
and so on.
Classical results can be found in \cite{Mu}--\cite{Bar2} 
and the references therein.
Hence the behaviors of their solutions have been studied 
analytically and thermostatistically in the literature 
\cite{FK}--\cite{Toscani}, 
just to name a few.
Further, these equations have been found to have close relations with 
optimal transport, Wasserstein metric and important inequalities 
in several branchs of mathematics such as 
functional analysis, probability theory or differential geometry
\cite{CT}--\cite{Toscani} and \cite{AGS}--\cite{Villani2}.
Many research results are summerized in recent monographs 
\cite{Frank05,Vaz06,Vaz07,AGS,Villani,Villani2}.

For a real number $q$, consider 
the {\em $q$-Gaussian probability density} function 
\cite{TLSM,ST} defined by
\begin{eqnarray}
G_q(x;\theta,\Theta):=
\exp_q\left(\theta^Tx +x^T \Theta  x - \psi(\theta, \Theta)\right), 
\label{qGauss} \\
\theta=(\theta^i) \in {\bf R}^n, 
\Theta=(\theta^{ij}) \in {\bf R}^{n \times n}, 
\nonumber
\end{eqnarray}
where $\exp_q t:=[1+(1-q)t]_+^{1/(1-q)}$, 
$\Theta$ is a real symmetric negative definite matrix
and $\psi(\theta,\Theta)$ is a normalizing constant.
The symbol $\cdot ^T$ denotes the transpose of a vector or matrix 
and $[a]_+$ for a real $a$ indicates $\max\{0,a\}$. 
Let ${\cal M}$ be the family of 
$q$-Gaussian densities specified by 
the parameters $(\theta,\Theta)$, i.e., 
\begin{equation}
 {\cal M}:=\left\{ \left. G_q(x;\theta, \Theta)
\right| \theta \in {\bf R}^n, \; 0>\Theta =\Theta^T \in {\bf R}^{n \times n} \right\}.
\label{qGmfd}
\end{equation}

The main purpose of the present paper is to study how solutions of 
the PME and NFPE behave relatively to 
the $q$-Gaussian family ${\cal M}$.
There are two major reasons for this novel viewpoint 
in the behavioral analysis.
First, ${\cal M}$ is proved to be an invariant manifold which 
all the solutions of the PME and NFPE asymptotically approach.
This implies that 
${\cal M}$ analogously plays a central role in the analysis to 
the {\em self-similar (Barenblatt-Pattle) solution} of the PME \cite{Vaz} or 
the asymptotically stable equilibria of the NFPE.
Hence, like the classical convergence analysis to the above two special 
solutions, we can expect to derive new interesting properties 
from this viewpoint.
Secondly ${\cal M}$ admits {\em information geometry} 
\cite{Amari,AN,Eguchi,MTKE}
compatible with the Legendre structure of {\em generalized thermostatistics} 
\cite{Naudts02,Naudts041}.
The geometry supplies to us several concepts 
such as projections or geodesics, which are useful tools to 
characterize a certain geometrical aspect of those solutions. 
They also give us clear physical interpretations, 
like evolutions of moments or the maximization of entropy.
Consequently, 
we derive several new and interesting geometrical properties and 
physical information of solutions to the PME and NFPE.
In mathematical statistics several families of multivariate distributions 
have been studied via information geometry \cite{Burb}--\cite{OC} 
because of their importance in many applications.
Further, Riemmanian geometries of the family of $q$-Gaussians 
${\cal M}$ are discussed with various Riemmanian metrics in \cite{Andai}.


In Section 2 we introduce and review Naudts' generalized thermostatistical 
theory \cite{Naudts02}--\cite{Naudts043} 
from the standpoint of information geometry, in particular, 
via recent work by Eguchi {\it et al.} \cite{Eguchi,MTKE}.
See also \cite{GD,David} for another context.
As a by-product the dualistic relation between two types of 
Bregman divergence is obtained, which clearly connects 
two formalisms proposed by Naudts and Eguchi {\it et al.}
Further, we define projections 
to the generalized exponential family 
and notions of geodesics, which are naturally induced from 
information geometric structure.
Among them the {\em m-projection} and {\em m-geodesic}, 
as well as the Bregman divergence, are our important tools 
to study the behavior.
In Section 3 
we demonstrate the main results on behaviors of solutions 
in terms of introduced geometric concepts on 
the $q$-Gaussian family ${\cal M}$.
We first prove that ${\cal M}$ is an invariant manifold for the PME and NFPE.
Next, utilizing the convenient property that the m-projection of a density 
to ${\cal M}$ conserves its first and second moment, 
we study the behavior of the solutions to the PME and NFPE.
Consequently, evolutions of the second moments and the convergence rate to 
the manifold are characterized by the divergence.
Further, the trajectory of the m-projection for a solution 
is proved to be an m-geodesic curve on ${\cal M}$.
Finally, we discuss a special geometric feature of the 
self-similar solution.

\section{Legendre structure on the generalized exponential family}
\label{geom}
\subsection{Generalized entropy and Bregman divergence}
Following \cite{Naudts02}--\cite{Naudts043}, we introduce generalized entropy 
and the Bregman divergence on the space of probability density functions. 
By bringing results derived from the {\em U-divergence} 
by Eguchi {\em et al.} \cite{Eguchi,MTKE} within a scope,
we show remarks that clarify the relation between their formalisms.

For a fixed strictly increasing and positive function $\phi(s)$ on 
$(0,\infty)$, define 
a {\em generalized logarithmic function} as follows:
\[
\ln_\phi(t):=\int_1^t \frac{1}{\phi(s)}ds, \quad t>0.  \;
\]
Note that $\ln_\phi$ is concave and strictly increasing and satisfies
$\ln_\phi(1)=0$. 
A {\em generalized exponential function} denoted by $\exp_\phi$ is defined as 
the inverse function of $\ln_\phi$, which can be extended on ${\bf R}$ by 
respectively putting 0 or $+\infty$ on the smaller or larger outside of 
the range $\ln_\phi$.
The function $\exp_\phi$ is confirmed 
to be strictly increasing and convex.
For two probability density functions $p(x)$ and $r(x)$, 
we define the {\em Bregman divergence} as follows:
\begin{equation}
 {\cal D}_\phi[p\|r]:=\int F_\phi(p(x))-F_\phi(r(x))-\ln_\phi r(x)(p(x)-r(x)) dx,
\label{DIVprimal}
\end{equation}
where $F_\phi(s)$ is defined for $s>0$ by
\begin{eqnarray}
 F_\phi(s):=\int_1^s \ln_\phi(t) dt, \quad 
F_\phi(0):=\lim_{s \rightarrow 0_+} F_\phi(s) <+\infty \hbox{ :assumed}.
\label{gl}
\end{eqnarray}
Note that $F_\phi$ is convex because $\ln_\phi $ is monotone
increasing. 
The divergence, if it exists, is positive except $p(x)=r(x)$ (a.e.).

Introduce a generalized entropy functional defined by
\begin{equation}
 {\cal I}_\phi[p]:= 
\int -F_\phi(p(x))+(1-p(x))F_\phi(0) dx,
\label{g_entropy}
\end{equation}
which is positive and concave with respect to $p$ 
because $F_\phi$ is convex and $F_\phi(1)=0$.
We omit the justification of the definition 
(\ref{g_entropy}) as the generalized entropy 
(see for \cite{Naudts02,Naudts041}) 
because it needs arguments 
for duality of the generalized logarithmic functions.

Here we should make two remarks.
First, defining a function on ${\bf R}$ by
\[
 U_\phi (t):=\int_0^t \exp _\phi u \; du
\]
and integrating the right-hand side of (\ref{gl}) by part, we have
\begin{eqnarray*}
 F_\phi(s)&=&s\ln_\phi s -\int_1^s t \frac{d \ln_\phi t}{dt} dt \\
 &=& s\ln_\phi s -\int_0^{\ln_\phi s} \exp_\phi u \; du \\
 &=& s\ln_\phi s - U_\phi(\ln_\phi s),
\end{eqnarray*}
or equivalently, 
\begin{equation}
	U_\phi(t)=t \exp_\phi t -F_\phi(\exp_\phi t).
\label{Leg}
\end{equation}
Thus, $U_\phi$ is regarded together with the relation (\ref{gl}), as 
the Legendre conjugate of $F_\phi $, and hence, is convex.
Then we have the dual expression of the divergence:
\begin{proposition}
The Bregman divergence (\ref{DIVprimal}) is expressed in the dual form by
\begin{equation}
 {\cal D}_\phi[p \| r]=\int U_\phi(\ln_\phi r(x))-U_\phi(\ln_\phi p(x))
	-p(\ln_\phi r(x)-\ln_\phi p(x)) dx.
\label{DIVdual}
\end{equation}
\end{proposition}
This form of the divergence is called the {\em U-divergence} \cite{Eguchi} 
and have been studied in the fields of statistics \cite{MTKE} because it is quite convenient in statistical inference 
from empirical data.
In this paper we mainly use this form to discuss geometry of 
generalized exponential family.

Second we see that the divergence is represented, 
using the entropy functional, by
\begin{eqnarray*}
 {\cal D}_\phi[p\|r]
&=&{\cal I}_\phi[r]-{\cal I}_\phi[p]-\int (p(x)-r(x))\ln_\phi r(x) dx \\
&=&\Phi_\phi[r]-{\cal I}_\phi[p]-\int p(x) \ln_\phi r(x) dx.
\end{eqnarray*}
Here the functional $\Phi_\phi [p]$ is defined by
\begin{eqnarray*}
 \Phi_\phi[p]:&=&\int p(x) \ln_\phi p(x) dx + {\cal I}_\phi[p] \\
&=&\int U_\phi(\ln_\phi p(x)) +(1-p(x))F_\phi(0) dx.
\end{eqnarray*}
As is seen below, $\Phi_\phi$ vanishes for the 
standard case $\phi(u)=u$.

\vspace{1em}
\noindent
{\bf Example ($q$-logarithmic and exponential functions)}: 
The results calculated in this example are used in section \ref{sec_main}.
Set $\phi(u)=u^q, q>0, q\not=1$, then we have the $q$-logarithmic and
exponential functions \cite{Naudts041,Naudts043}:
\[
 \ln_\phi t=\ln_q t :=(t^{1-q}-1)/(1-q), \quad
 \exp_\phi t=\exp_q t :=[1+(1-q)t]_+^{1/(1-q)},
\]
where $[x]_+=\max\{x,0\}$ for $x \in {\bf R}$. 
Note that when $q \rightarrow 1$, they recover natural logarithmic and 
exponential functions.
If $2-q>0$, we have $F_\phi(s)$ and the constant $F_\phi(0)$, respectively,
\[
 F_\phi(s)=\int_1^s \frac{t^{1-q}-1}{1-q}dt 
= \frac{1}{1-q}\left( \frac{s^{2-q}}{2-q}-s \right)+\frac{1}{2-q}, \quad 
F_\phi(0)=\frac{1}{2-q}.
\]
Consequently it reproduces 
the corresponding generalized entropy ${\cal I}_\phi$: 
\begin{equation}
 {\cal I}_\phi[p] 
= \frac{1}{2-q}\int \frac{p(x)^{2-q}-p(x)}{q-1}dx 
=   \frac{1}{2-q} {\cal S}_{2-q}[p],
\label{dualent}
\end{equation}
where is ${\cal S}_q$ is called the {\em Tsallis entropy} \cite{Tsallis1} defined by
\[
{\cal S}_q[p]= \int \frac{p(x)^q-p(x)}{1-q} dx. 
\]
Note that ${\cal S}_{2-q}$ is represented by
\[
  {\cal S}_{2-q}[p]=- \int p(x)\ln_q p(x) dx
	= \int p(x)\ln_{2-q}\left( \frac{1}{p(x)} \right) dx.
\]
Consider the case $1-q>0$, which is used in section \ref{sec_main}, 
then for $s \ge -1/(1-q)$ we have 
\[
 U_\phi(s) = \frac{1}{2-q} \{ 1+(1-q)s \}^{(2-q)/(1-q)}-\frac{1}{2-q}. 
\]
Since $\ln_q t \ge -1/(1-q)$ for $t \ge 0$, we see that
\[
	U_\phi(\ln_\phi s)=\frac{1}{2-q}(s^{2-q}-1), \; s \ge 0.
\]
Then the corresponding Bregman divergence of the form (\ref{DIVdual}) is
\begin{equation}
	{\cal D}_\phi[p\|r]=\int \frac{r(x)^{2-q}
	-p(x)^{2-q}}{2-q}-p(x)\frac{r(x)^{1-q}-p(x)^{1-q}}{1-q}dx, 
\end{equation}
which is called the {\em $\beta$-divergence} and 
applied to robust estimation in statistics or machine learning 
\cite{ME,HE,TE,MTKE}.
Finally, we have
\[
 \Phi_\phi[p]=\frac{1}{2-q} \int p(x)^{2-q} -p(x) dx.
\]
This functional disappears when $q \rightarrow 1$.

\subsection{Generalized exponential family and its geometry}

Let us consider the following finite dimensional statistical model 
called {\em the generalized exponential family} \cite{Naudts043,GD} 
or {\em $U$-statistical model} \cite{Eguchi}, 
which is defined by
\[
 \mathcal{M}_\phi := 
\{p_\theta(x)=\exp_\phi(\theta^T h(x)-\psi_\phi(\theta)) \; | \; \theta \in 
\Omega \subset {\bf R}^d \} \subset L^1({\bf R}^n).
\]
Here $h(x)=(h_i(x)), i=1,\cdots,d$ is a certain vector-valued function 
and $\psi_\phi(\theta)$ is a normalizing factor of $p_\theta$, i.e., 
\begin{equation}
 \int \exp_\phi(\theta^T h(x)-\psi_\phi(\theta)) dx =1.
\label{normality}
\end{equation}
On the open domain $\Omega$ we assume that the regularity condition holds, 
i.e., all the formal calculus 
necessary below such as convergence of integrations, differentiability 
of the map from $\theta$ to $p_\theta$ and so on are valid, 
so that we can regard ${\cal M}_\phi$ 
as a differentiable manifold.
Since the parameter $\theta=(\theta^i)$ specifies a density function 
in ${\cal M}_\phi$, it plays a role of the coordinate system 
for ${\cal M}_\phi$.
Differentiating (\ref{normality}) by $\theta^i$ 
(we denote basis tangent vectors by 
$\partial_i:=\partial/\partial \theta^i$), we have
\begin{equation}
	\int \exp'_\phi(\theta^Th(x)-\psi_\phi(\theta))
			(h_i(x)-\partial_i \psi_\phi(\theta)) dx =0,
\label{from_normal}
\end{equation}
which will be used later.

One of the simplest ways to define information geometric structure 
\cite{Amari,AN}
on ${\cal M}_\phi$, which is natural with 
the generalized entropy ${\cal I}_\phi$, is 
invoking the following {\em potential function}:
\begin{eqnarray*}
 \Psi_\phi(\theta):= \Phi_\phi[p_\theta]+\psi_\phi(\theta) 
= \int U_\phi(\ln_\phi p_\theta)+(1-p_\theta(x))F_\phi(0) dx 
+\psi_\phi(\theta).
\end{eqnarray*}
Note that in the standard case $\phi(u)=u$, we have 
$\Psi_\phi(\theta)=\psi_\phi(\theta)=\log \int \exp \theta^Th(x) dx$ 
because $\Phi_\phi$ vanishes 
as is seen in the example of the previous section.
It follows from the relation $\exp_\phi=U'_\phi$ that
\begin{equation}
 \eta_i(\theta):=\partial_i \Psi_\phi(\theta)= \int h_i(x)p_\theta(x)dx
=\mathbf{E}_{p_\theta}[h_i(x)],
\label{expectation}
\end{equation}
where we denote by ${\bf E}_p[\cdot]$ the expectation operator 
for the density $p$.
Using (\ref{from_normal}) we have
\begin{eqnarray}
 \partial_i \partial_j \Psi_\phi(\theta)
	&=& \int  h_i(x) \exp'_\phi(\theta^T h(x)-\psi_\phi(\theta))(h_j(x)
		-\partial_j \psi_\phi(\theta))dx \nonumber \\
&=& \int \tilde h_i(x) 
	\exp'_\phi(\theta^T h(x)-\psi_\phi(\theta)) 
	\tilde h_j(x) dx, 
\label{Hesse}
\end{eqnarray}
where $\tilde h_i(x):=h_i(x)-\partial_i \psi_\phi(\theta)$.
Thus, the Hesse matrix of $\Psi_\phi(\theta)$ is expressed by
\begin{eqnarray*}
	(\partial_i \partial_j \Psi_\phi)=
	\int \tilde h(x) \tilde h^T(x) 
		\exp'_\phi(\theta^T h(x)-\psi_\phi(\theta)) dx, \\
	\mbox{where } \tilde h(x)=
		\left( \tilde h_1(x) \; \cdots \tilde h_n(x) \right)^T,
\end{eqnarray*}
and we see that it is positive semidefinite 
because $\exp'_\phi$ is nonnegative, 
and hence, $\Psi_\phi$ is a convex function of $\theta$.
In the sequel, we assume that 
$(\partial_i \partial_j \Psi_\phi)=(\partial \eta_j/\partial \theta^i)$ 
is positive definite for $\forall \theta \in \Omega$.
Hence, $\eta=(\eta_j)$ is locally bijective to $\theta$ and 
we call $(\eta_i)$ the {\em expectation coordinate system} 
for ${\cal M}_\phi$.
By the relation (\ref{expectation}) the Legendre conjugate of 
$\Psi_\phi(\theta)$ is the sign-reversed generalized entropy 
of $p_\theta \in {\cal M}_\phi$, i.e,
\begin{eqnarray}
 \Psi^*_\phi(\eta)&:=& \theta^T \eta - \Psi_\phi(\theta) 
= \int p_\theta \log_\phi p_\theta  - U_\phi(\ln_\phi p_\theta) -(1-p_\theta)F_\phi(0) dx \nonumber \\
&=& -{\cal I}_\phi[p_\theta].
\label{Massieu}
\end{eqnarray}
Hence, $\Psi_\phi$ can be physically interpreted as the 
{\em generalized Massieu potential} \cite{Callen,WS05}, 
and hence our Riemmanian metric $(\partial_i \partial_j \Psi_\phi)
=(\partial \eta_j/\partial \theta^i)$ defined below is regarded as a 
{\em susceptance} matrix.


As a {\em Riemannian metric} $g=(g_{ij})$ on ${\cal M}_\phi$, 
which is an inner product for tangent vectors, 
we use the Hesse matrix of $\Psi_\phi$.
Note that we can alternatively express (\ref{Hesse}) as
\[
 g_{ij}(\theta)=g(\partial_i,\partial_j):= \partial_i \partial_j \Psi_\phi
=\int \partial_i p_\theta(x) \partial_j
 \ln_\phi p_\theta(x) dx.
\]
For an explict formula of $g$ in the case of 
$q$-Gaussian densities ${\cal M}$, see, for example, \cite{Andai}.
Further we define {\em generalized mixture connection} 
$\nabla^{({\rm m})}$ and {\em exponential connection} 
$\nabla^{({\rm e})}$  
by their components 
\[
 \Gamma^{({\rm m})}_{ij,k} (\theta) 
	=g(\nabla^{({\rm m})}_{\partial_i} \partial_j,\partial_k)
	:= \int \partial_i \partial_j p_\theta(x) 
		\partial_k \ln_\phi p_\theta(x) dx,
\]
\begin{equation}
 \Gamma^{({\rm e})}_{ij,k} (\theta) 
	=g(\nabla^{({\rm e})}_{\partial_i} \partial_j,\partial_k)
	:= \int \partial_k  p_\theta(x) 
		\partial_i \partial_j \ln_\phi p_\theta(x) dx.
\label{dual_connex} 
\end{equation}
Then the {\em duality relation of the connections} \cite{Amari,AN}
$
	\partial_i g_{jk}=\Gamma^{({\rm m})}_{ij,k}+\Gamma^{({\rm e})}_{ik,j}
$
holds.
From the definition we see the coordinate system $(\theta^i)$ is special 
in the sense that 
$\Gamma^{({\rm e})}_{ij,k}$ actually vanishes, i.e., 
\[
\Gamma^{({\rm e})}_{ij,k} (\theta) = - \int \partial_k  p_\theta(x)
 \partial_i \partial_j \psi_\phi(\theta) dx = - \partial_i \partial_j
 \psi_\phi(\theta) \partial_k \int p_\theta(x)dx =0.
\]
Hence, $\nabla^{({\rm e})}$ is a flat connection 
on ${\cal M}_\phi$ and 
$(\theta^i)$ is an {\em affine coordinates} with respect to 
$\nabla^{({\rm e})}$, in other words, each $\partial_i$ is parallel 
with respect to $\nabla^{({\rm e})}$. 
Note that we have $
	\Gamma^{({\rm m})}_{ij,k} 
	=\partial_i \partial_j \partial_k \Psi_\phi
$.
Similarly, we can show that $\nabla^{({\rm m})}$ is also flat 
on ${\cal M}_\phi$ 
and $(\eta_j)$ is affine with respect to 
$\nabla^{({\rm m})}$
via formal argument based on the duality \cite{Amari,AN}.

Thus, we have obtained {\em dually flat} \cite{Amari,AN} structure 
$(g,\nabla^{({\rm m})}, \nabla^{({\rm e})})$  
on ${\cal M}_\phi$ introduced by the derivatives of $\Psi_\phi$.
Note that if $\phi(u)=u$, then 
${\cal M}_\phi$, $g$, $\nabla^{({\rm m})}$ and $\nabla^{({\rm e})}$ 
respectively reduce to the {\em exponential family}, 
the {\em Fisher information matrix}, 
the usual {\em mixture and exponential connections}.
We shall find in the sequel that the structure offers useful tools to us for 
not only the statistical 
inference but also the analysis of the PME or NFPE.

The next result immediately follows from the fact that 
the coordinates are affine with respect to the flat connections, 
and is frequently used in this paper.
\begin{proposition}
\label{prop_geod}
Let ${\cal C}$ be a one-dimensional 
submanifold in ${\cal M}_\phi$.
Each coordinate $\theta^i$ of ${\cal C}$ is a linear function of 
a common scalar variable, i.e., ${\cal C}$ is expressed as a 
straight line in the coordinate system $\theta=(\theta^i)$, 
if and only if ${\cal C}$ coincides with a $\nabla^{({\rm e})}$-geodesic 
({\em e-geodesic}, in short) curve.
Similarly, ${\cal C}$ is expressed as a 
straight line in the coordinate system $\eta=(\eta_i)$ if and only if
${\cal C}$ coincides with a 
$\nabla^{({\rm m})}$-geodesic ({\em m-geodesic}) curve.
\end{proposition}

In the rest of this subsection we assume that $\ln_\phi \circ \exp_\phi (t) =t$ holds for all $t \in {\bf R}$.
For densities $p_\theta(x) \in {\cal M}_\phi$ and $p(x)$, 
we have
\begin{eqnarray}
 {\cal D}_\phi[p \| p_{\theta}]&=&\int U_\phi(\ln_\phi p_\theta(x))-U_\phi(\ln_\phi 
 p(x)) \nonumber \\
&& \qquad -p(x)\left[\theta^T h(x) -\psi_\phi(\theta) -\ln_\phi p(x)\right] dx \nonumber \\
 &=& \Psi_\phi(\theta) + \int F_\phi(p(x)) - (1-p(x))F_\phi(0) dx - \theta^T {\bf E}_p[h(x)] 
\nonumber \\
 &=&\Psi_\phi(\theta)-{\cal I}_\phi[p]-\theta^T {\bf E}_p[h(x)].
\label{div_der}
\end{eqnarray}
Further, for $p_{\theta_1}$ and $p_{\theta_2}$ 
both in ${\cal M}_\phi$, the Bregman divergence is represented by
\begin{eqnarray}
  {\cal D}_\phi[p_{\theta_1} \| p_{\theta_2}] 
	&=& \Psi_\phi(\theta_2) + \Psi_\phi^*(\eta_1)  - \eta_1^T \theta_2 
\nonumber \\
&=& \Psi_\phi(\theta_2)-\Psi_\phi(\theta_1)-\eta_1^T(\theta_2-\theta_1),
\label{div_der2}
\end{eqnarray}
where $\eta_1$ and $\eta_2$ are the expectation coordinates for  
$p_{\theta_1}$ and $p_{\theta_2}$, respectively.
Using the above, we introduce the notion of {\em m-projection}, which is 
geometrically related to the m-geodesic and orthogonality \cite{Amari,AN}, 
and prove important properties essential in the behavioral analysis.

\begin{definition}
Let $p(x)$ be a given density. 
If there exists a minimizing density $\hat p_\theta(x)$ 
for the variational problem 
$\min_{p_\theta \in {\cal M}_\phi} {\cal D}_\phi[p \| p_\theta]$,
or equivalently, a minimizing parameter $\hat \theta$ for the problem 
$\min_{\theta \in \Omega}{\cal D}_\phi[p \| p_\theta]$ 
exists, we call $\hat p_\theta(x)=p_{\hat \theta}(x)$ 
the {\em m-projection of $p(x)$ to ${\cal M}_\phi$}.
\end{definition}

\begin{proposition}
\label{prop_proj}
Let $\hat p_\theta \in {\cal M}_\phi$ be the m-projection of $p$.
Then the following properties hold:
\begin{description}
\item[i)] The m-projection conserves the expectation of $h(x)$, i.e., 
${\bf E}_p[h(x)]={\bf E}_{\hat p_\theta}[h(x)]$,
\item[ii)] The following triangular equality holds:
$
	{\cal D}_\phi[p \| p_\theta]={\cal D}_\phi[p \| \hat p_\theta] 
	+ {\cal D}_\phi[\hat p_\theta \|p_\theta]
$ for all $p_\theta \in {\cal M_\phi}$.
\end{description}
\end{proposition}
Proof) 
Consider the optimality condition for the convex optimization problem 
$
	\min_{p_\theta \in {\cal M}} {\cal D}_\phi[p \| p_\theta].
$
Since the second term of the right-hand side of (\ref{div_der}) is constant, 
we have 
\begin{equation}
	\frac{\partial \Psi_\phi}{\partial \theta^i}(\theta) 
		- {\bf E}_{p}[h_i(x)]=0, \quad i=1,\cdots,d.
\label{optimal}
\end{equation}
Let $\hat \theta$ be the solution for the above optimality condition, i.e., 
$\hat p_\theta=p_{\hat \theta}$.
Then, from (\ref{expectation}) and (\ref{optimal}), 
the statement i) holds.

For the statement ii), we use (\ref{div_der}) and (\ref{div_der2}).
Since the statement i) implies that  
$\hat \eta:={\bf E}_{\hat p_\theta}[h(x)]={\bf E}_{p}[h(x)]$, and 
the identity $\Psi_\phi(\hat \theta)+\Psi_\phi^*(\hat \eta)-\hat \theta^T \hat \eta=0$ holds from (\ref{Massieu}), 
we can modify the right-hand side of the triangular equality as
\begin{eqnarray}
	\Psi_\phi(\hat \theta)-{\cal I}_\phi[p]-\hat \theta^T {\bf E}_{p}[h(x)]
	+\Psi_\phi(\theta)+\Psi^*_\phi(\hat \eta)-\theta^T \hat \eta 
	\nonumber \\
	=\Psi_\phi(\theta)-{\cal I}_\phi[p]-\theta^T {\bf E}_{p}[h(x)],
\end{eqnarray}
which is equal to ${\cal D}_\phi[p \| p_\theta]$.
\hfill Q.E.D.

\begin{remark}
\label{rem_proj}
From the statement i) the m-projection $\hat p_\theta$ 
is characterized as the density in 
${\cal M}_\phi$ with the expectation coordinate equal to ${\bf E}_p[h(x)]$.
In other words, for any $p$ satisfying ${\bf E}_p[h(x)]
={\bf E}_{\hat p_\theta}[h(x)]=\hat \eta$, we have
\begin{eqnarray*}
	{\cal D}_\phi[p \| \hat p_{\theta}]&=&\Psi_\phi(\hat \theta)-{\cal I}_\phi[p]
	-\hat \theta^T {\bf E}_p[h(x)] \\
	&=&\Psi_\phi(\hat \theta)-\hat \theta^T \hat \eta-{\cal I}_\phi[p]
	={\cal I}_\phi[\hat p_\theta]-{\cal I}_\phi[p] \ge 0.
\end{eqnarray*}
Thus, $\hat p_\theta$ achieves the {\em maximum entropy} among densities $p$ 
with the equal expectation of $h(x)$.
\end{remark}

\section{Several geometric properties of the porous medium and the associated Fokker-Planck equation}
\label{sec_main}
\subsection{Preliminaries}
In this section, 
we study the Cauchy problems of the PME (\ref{PME}) and the NFPE (\ref{DNFPE}) 
from a viewpoint of
information geometry on the $q$-Gaussian family ${\cal M}$ 
given in (\ref{qGmfd}).
In other words, we apply the general argument in the previous section 
to the case 
where $\phi(u)=u^q, q>0$ (cf. the example in section 2) 
and $h(x)$ is quadratic, i.e., 
$\theta^T h(x)$ is actually $\Theta \cdot (xx^T)+ \theta \cdot x$.
Here, $\cdot$ used for matrices $A$ and $B$ 
denotes their inner product, i.e., $A \cdot B={\rm trace}(A^TB)$.

In the sequel we fix the relation between the exponents of the PME 
and the parameter of the $q$-exponential function by $m=2-q$.
Hence, we consider the case $1<m<2$, or equivalently, $0<q<1$.
For the brevity, we omit the subscripts $\phi$.
Further, for 
$G_q(x;\theta,\Theta) \in {\cal M}$ and an arbitrary density $p(x)$, 
we modify the divergence (9) as
\begin{eqnarray*}
	{\cal D}[p(x) \| G_q(x;\theta,\Theta)]&:=& 
	\int U(\ln_q G_q(x;\theta,\Theta))-U(\ln_q p(x))  \\
	&& \; -p(x)[\theta^Tx+x^T \Theta x-\psi(\theta,\Theta)- \ln_q p(x)] dx.
\end{eqnarray*}
The modified divergence also satisfies positivity, convexity in $\theta$ 
and Proposition 3.
In the NFPE (\ref{DNFPE}) we can always choose $\beta$ as an arbitrary 
constant by a suitable linear scaling of $\tau$ and $D$.
Hence, we set $\beta$ and 
introduce another constant $\alpha$ for notational simplicity as follows:
\[
	\beta:=\frac{1}{n(m-1)+2}, \quad \alpha:=n \beta.
\]

For the $q$-Gaussian family ${\cal M}$, we can regard 
$(\theta,\Theta)$ as the canonical coordinates.
On the other hand, the expectation coordinates are 
nothing but the first moment vector and 
second moment matrix $(\eta,H)$ defined by
\[
	\eta=\int xG_q(x;\theta,\Theta)dx, \quad 
	H=\int xx^T G_q(x;\theta,\Theta)dx.
\]
Note that the dimension of ${\cal M}$ is $N:=n+n(n+1)/2=n(n+3)/2$.

We assume that $u(x,0)$ and $p(x,0)$, which respectively denote the 
initial data of the PME and NFPE, are
nonnegative and integrable function with finite second moments.
Under these assumptions, it is proved 
that there exists a unique nonnegative weak solution 
if $m>0$, and that 
the mass $M=\int u(x,t)dx$ is conserved for all $t>0$ 
if $m \ge (n-2)/n$. 
See \cite{Vaz} for details and additional properties.
When we consider the solutions, we restrict their initial masses 
to be normalized to one without loss of generalities.

In this subsection we show three fundamental facts, of which the 
last one might be new.
First of all, we review how the solutions of the PME and NFPE relate 
each other (Cf. \cite{Vaz,Toscani}).
Because of this fact the properties of the solution of the PME (\ref{PME}) 
are important to investigate those of NFPE (\ref{DNFPE}) and vise versa.


\begin{proposition}
\label{prop_transfer}
Let $u(x,t)$ be a solution of the PME (\ref{PME}) with initial data 
$
	u(x,0)=u_0(x) \in L^1({\bf R}^n),
$
Define 
\[
	p(z,\tau):=(t+1)^\alpha u(x,t), \quad z:=(t+1)^{-\beta}Rx , \; \tau:=\ln (t+1),
\]
then $p(z,\tau)$ is a solution of (\ref{DNFPE}) with $\nabla=\nabla_z$, 
$D=RR^T$ 
and initial data
$
	p(z,0)= u_0(R^{-1}z).
$
\end{proposition}
Proof: It is known \cite{Vaz} that 
\begin{equation}
 p(y,\tau):=(t+1)^\alpha u(x,t), \quad y:=x (t+1)^{-\beta}, \; \tau:=\ln (t+1)
\label{PMEtoFPE}
\end{equation}
is a solution of the following NFPE:
\begin{equation}
 \frac{\partial p}{\partial \tau}
	= \nabla_y \cdot (\beta y p + \nabla_y p^m),
\label{NFPE} 
\end{equation}
where $\nabla_y=(\partial/\partial y^1 \; \cdots \; \partial/\partial y^n)^T$.
Since it holds, for the transformation $z=Ry$, that
\[
\nabla_z=
\left( 
   \frac{\partial}{\partial z^1} \cdots
\displaystyle  \frac{\partial}{\partial z^n}
\right)^T
=R^{-T} \nabla_y,
\]
we see that $p(y,\tau)$ is the solution of (\ref{NFPE}) if and only if 
$p(R^{-1}z,\tau)$ is the solution of 
\[
 \frac{\partial p(R^{-1}z,\tau)}{\partial \tau}=  
\nabla_z \cdot \left(\beta z p(R^{-1}z,\tau)+D\nabla_z p(R^{-1}z,\tau)^m
\right).
\]
Note that the drift vector is invariant. 
Thus, the statement follows.
\hfill Q.E.D.

\vspace{1em}
Next, the equilibrium density for the NFPE (\ref{DNFPE}) is solved 
using generalized thermostatistical concept and Lyapunov approach.
To analyze the behavior of (\ref{DNFPE}) 
let us define a {\em generalized free energy}:
\begin{eqnarray}
 {\cal F}[p]
	&:=& \int \frac{\beta}{2m} x^T D^{-1} x p(x) dx -{\cal I} [p] 
		\nonumber \\
	&=&\frac{1}{m} 
	\int \frac{\beta}{2} x^T D^{-1} x p(x) + p(x) \ln_q p(x) dx.
\end{eqnarray}
This type of functional was first introduced in \cite{Newman,Ralston} 
and developed by many researchers \cite{Frank02,Vaz,Toscani} 
to discuss convergence of the PME and NFPE. 
Note that when $n=1$, it reduces to $U/D-S_{2-q}$ up to constant, with 
the average energy $U={\bf E}_p[\beta x^2/2]$ for a drift vector $\beta x$.
Hence, the diffusion coefficient $D$ can be interpreted as the temperature 
in the thermodynamical argument.

%

Note that (\ref{DNFPE}) can be rewritten as
\[
   \frac{\partial p(x,\tau)}{\partial \tau}
	= (R\nabla) \cdot \left[ \beta R^{-1}x p(x,\tau) 
	+ R \nabla p(x,\tau)^{2-q} \right]
\]
using a symmetric matrix $R$ satisfying $R^2=D$.
Recalling the relation $q=2-m$, we have
\[
	\frac{\delta {\cal F}}{\delta p} 
	= \frac{1}{m} \left\{ \frac{\beta}{2}(R^{-1}x)\cdot(R^{-1}x)
		-\frac{mp(x,\tau)^{m-1}-1}{1-m} \right\}.
\]
Hence, it holds that
\begin{eqnarray}
\frac{d{\cal F}[p(x,\tau)]}{d\tau} &=& \int \frac{\delta {\cal F}}{\delta p} 
	\frac{\partial p}{\partial \tau} dx 
	= \int \frac{\delta {\cal F}}{\delta p} (R \nabla) \cdot 
		(\beta R^{-1}x p+R\nabla p^{m}) dx \nonumber \\
&=& - \int \left((R \nabla) \frac{\delta {\cal F}}{\delta p} \right) \cdot 
(\beta R^{-1}x p+R\nabla p^{m}) dx \nonumber \\
&=& - \frac{1}{m}\int p \| \beta R^{-1} x + m p^{m-2}R \nabla p \|^2 dx \le 0.
\label{Lyapunov}
\end{eqnarray}
Thus, ${\cal F}[p(x,\tau)]$ serves as a Lyapunov functional for 
(\ref{DNFPE}) and
the equilibrium density $p_\infty(x)$ is  
determined from (\ref{Lyapunov}) as a $q$-Gaussian:
\begin{eqnarray}
	p_\infty(x)=G_q(x;0,\Theta_\infty)&=& 
	\exp_q(x^T \Theta_\infty x-\psi(0, \Theta_\infty)), \nonumber
	\\
&=& \exp_q (\Theta_\infty \cdot (xx^T)-\psi(0,\Theta_\infty)),
\label{eq_density}
\end{eqnarray}
where the parameters are given by
\[
	\theta_\infty=0, \quad \Theta_\infty=-\frac{\beta}{2m} D^{-1}.
\]


From (16) the generalized Massieu potential 
on the $q$-Gaussian family ${\cal M}$ is represented by
\[
	\Psi(\theta, \Theta)=\theta^T {\bf E}_{G_q{(\theta,\Theta)}}[x]
		+ \Theta \cdot {\bf E}_{G_q{(\theta,\Theta)}}[xx^T]  
		+ {\cal I}[G_q{(\theta,\Theta)}].
\]
Since the generalized free energy is written as
\[
	{\cal F}[p]=-\Theta_\infty \cdot {\bf E}_{p}[xx^T]-{\cal I}[p],
\]
we can express the difference of the free energies at  
$p_\infty(x) \in {\cal M}$ and $p(x)$ 
by the divergence via (18):
\begin{eqnarray*}
 {\cal D}[p||p_\infty] &=&\Psi(0,\Theta_\infty)
	-{\cal I}[p] 
	- \Theta_\infty \cdot {\bf E}_p[xx^T]  \\
	&=& {\cal F}[p]-{\cal F}[p_\infty].
\end{eqnarray*}
Thus, the minimization of ${\cal F}[\cdot]$ 
is equivalent to that of ${\cal D}[\cdot \|p_\infty]$, 
which can be interpreted as the {\em H-theorem} in 
statistical physics.

\vspace{1em}
Finally, we show the $q$-Gaussian family ${\cal M}$ is 
invariant and attracting for the PME and NFPE, i.e., 
a solution belongs to ${\cal M}$ for all future time if 
its initial density is in ${\cal M}$, and converges to ${\cal M}$ 
otherwise.
Hence, together with the previous fact on the equilibrium, 
analysis with respect to ${\cal M}$ is expected to 
add basic knowledge about the behaviors of the solutions.

\begin{proposition}
The $q$-Gaussian family ${\cal M}$ is an invariant and attracting 
manifold for the PME and NFPE.
\end{proposition}
Proof)  
We prove that $\Delta G_q(x;\theta,\Theta)^m$ belongs to the tangent space of 
${\cal M}$ at each $G_q(x;\theta,\Theta)$.
For the $q$-Gaussian density $G_q(x;\theta,\Theta)$ defined by 
(\ref{qGauss}), we see that $\Delta G_q(x;\theta,\Theta)^m$ is of the form 
\[
	\Delta G_q(x;\theta,\Theta)^m=\left\{
		\begin{array}{cl}
		Q(x;\theta,\Theta)G_q(x;\theta,\Theta)^{2-m}, & 
			x \in {\rm supp} G_q(x;\theta,\Theta), \\
		0, & x \not\in {\rm supp}G_q(x;\theta,\Theta),
		\end{array}
		\right. 
\]
with a certain quadratic function of $x$, i.e, 
$Q(x;\theta,\Theta)=x^TAx+b^Tx+c$, 
where the coefficients $A=(a_{ij}), b=(b_i)$ and $c$ depend on 
$\theta=(\theta^i)$ and $\Theta=(\theta^{ij})$. 
Note that it holds that
\[
	\int \Delta G_q(x;\theta,\Theta)^m d \mu =0
	\quad \forall \theta, \; \forall \Theta
\]
from the mass conservation property of the PME.
Hence, $A,b$ and $c$ have a linear constraint and 
the scalar coefficient $c$ is determined by $A$ and $b$.
On the other hand, the natural tangent basis vectors of ${\cal M}$ are 
calculated by
\begin{eqnarray*}
	\frac{\partial G_q}{\partial \theta^{i}} (x;\theta,\Theta)
	&=& \left\{
	\begin{array}{cl}
	\displaystyle
	\left(x_i + \frac{\partial \psi}{\partial \theta^{i}}\right)
	G_q(x;\theta,\Theta)^{2-m}, & 
		x \in {\rm supp} G_q(x;\theta,\Theta), \\
	0, & x \not\in {\rm supp}G_q(x;\theta,\Theta),
	\end{array}
	\right. \\
	\frac{\partial G_q}{\partial \theta^{ij}} (x;\theta,\Theta)
	&=& \left\{
	\begin{array}{cl}
	\displaystyle
	\left((2-\delta_{ij})x_ix_j +\frac{\partial \psi}{\partial \theta^{ij}}
		\right)
	G_q(x;\theta,\Theta)^{2-m}, & 
		x \in {\rm supp} G_q(x;\theta,\Theta), \\
	0, & x \not\in {\rm supp}G_q(x;\theta,\Theta),
	\end{array}
	\right.  
\end{eqnarray*}
for $i,j=1,\cdots,n$ with $i \le j$.
Here, $\delta_{ij}$ is the Kronecker's delta.
From the definition of ${\cal M}$ they also conserve the mass, i.e.,
\[
	\int \frac{\partial G_q}{\partial \theta^{i}}(x;\theta,\Theta) dx =0,		\quad 
	\int \frac{\partial G_q}{\partial \theta^{ij}}(x;\theta,\Theta)dx =0,
	\quad \forall \theta, \; \forall \Theta.
\]
Consider the following linear combination of these tangent vectors 
with the coefficients $a_{ij}$ and $b_i$:
\[
	v:=\sum_{i,j} a_{ij} \frac{\partial G_q}{\partial \theta^{ij}}(x;\theta,\Theta)
		+ \sum_i b_i \frac{\partial G_q}{\partial \theta^{i}}(x;\theta,\Theta).
\]
Then $v$ is equal to $\Delta G_q(x;\theta,\Theta)^m$ 
because the both satisfy the mass conservation constraint and 
the remaining coefficient $c$ should be consequently represented by 
\[
	c=\sum_{i,j} a_{ij}\frac{\partial \psi}{\partial \theta^{ij}}				+ \sum_i b_i \frac{\partial \psi}{\partial \theta^{i}}
		\quad \forall \theta, \; \forall \Theta.
\]
Thus, $\Delta G_q(x;\theta,\Theta)^m$ belongs to the tangent space of 
${\cal M}$ at $G_q(x;\theta,\Theta)$.
The attractivity is straightforward from the well-known fact 
that all the solutions 
of the PME asymptotically converge to the self-similar (Barenblatt-Pattle) 
solution $u^{\rm BP}(x,t)$ lying on ${\cal M}$. (Cf. subsection \ref{SSsol}.)
The invariance and attractivity of ${\cal M}$ for the NFPE follows 
from this result and the transformation in Proposition \ref{prop_transfer}.
\hfill Q.E.D.

\subsection{Trajectories of m-projections}
First we study the behavior of a solution $u(x,t)$ of the PME in terms of its 
m-projection to ${\cal M}$ denoted by $\hat u(x,t)$.
Owing to the properties of the divergence described in section \ref{geom},
this is equivalent to consider the first and second moments of $u(x,t)$.
Let $\eta^{\rm PM}=(\eta^{\rm PM}_i)$ and $H^{\rm PM}=(\eta^{\rm PM}_{ij})$ 
be, respectively, the first moment vector and 
the second moment matrix of $u$: 
\[
	\eta^{\rm PM}_{i}(t):=\mathbf{E}_u[x_i], 
	\quad \eta^{\rm PM}_{ij}(t):=\mathbf{E}_u[x_ix_j].
\]

\begin{theorem}
\label{thm1}
Consider solutions of the PME 
with the common initial first and second moments.
Then their m-projections to ${\cal M}$ evolve monotonically along 
with the common m-geodesic curve that starts from the density 
determined by the initial moments.
\end{theorem}
Proof) 
Differentiating $\eta^{\rm PM}_{ij}$ by $t$, we have
\begin{eqnarray*}
 \dot \eta^{\rm PM}_{ij}&=&\int \frac{\partial u}{\partial t} x_i x_j dx 
= \int \Delta u^m x_i x_j dx 
= - \int \nabla u^m  \cdot \nabla \left( x_ix_j \right) dx \nonumber \\
&=& \int u^m \Delta \left( x_ix_j \right) dx 
= 2\delta_{ij} \int u^m dx. 
\end{eqnarray*}
Hence, the second moment evolves as follows:
\[
 \eta^{\rm PM}_{ij}(t)=\eta^{\rm PM}_{ij}(0)+ \delta_{ij} \sigma_u^{\rm PM}(t), 
\quad \sigma_u^{\rm PM}(t):= 2 \int_0^t dt' \int u(x,t')^m dx.
\]
Note that $\sigma_u^{\rm PM}(t)$ is positive and monotone increasing on $t>0$.
By similar argument we see that $\dot \eta^{\rm PM}=0$, i.e., 
the first moment vector is invariant.
Thus, the evolution of 
$(\eta^{{\rm PM}}(t),H^{{\rm PM}}(t))$ for every solution $u(x,t)$
is represented as a straight line.
Recalling that the m-projection conserves these moments 
from Proposition \ref{prop_proj}, 
we see that $(\eta^{{\rm PM}}(t),H^{{\rm PM}}(t))$ 
is just the expectation coordinates of $\hat u(x,t)$.
Thus, the trajectory of the m-projection of $u(x,t)$ is 
an m-geodesic curve by Proposition \ref{prop_geod}.
\hfill Q.E.D.
\begin{remark}
\label{rem_thm}
i) From the argument for the NFPE, we will see that $\sigma_u^{\rm PM}(t)=O(t^{2 \beta})$ as 
$t \rightarrow \infty$. 
\par \noindent
ii) Theorem implies 
that the trajectories of m-projections on ${\cal M}$ 
for all the PME solutions $u(x,t)$ 
are parallelized in the expectation coordinates, i.e., 
\begin{eqnarray}
	\eta^{\rm PM}(t)&=&\eta^{\rm PM}(0), \\
	H^{\rm PM}(t)&=&H^{\rm PM}(0)+\sigma_u^{\rm PM}(t)I,
\label{sol_mom}
\end{eqnarray}
where $I$ is the $n$ by $n$ identity matrix.
In other words, the PME has the following $N(={\rm dim}{\cal M})$ 
{\em constants of motion} including the mass $M$:
\begin{eqnarray*}
	J_0 (=M):= \int u(x,t)dx, \\
	J_i (=\eta^{\rm PM}_i) := \int x_i u(x,t)dx, \quad i=1,\cdots,n, \\ 
	J_{ij} (=\eta^{\rm PM}_{ij}) := \int x_ix_j u(x,t)dx, 
		\quad i=1,\cdots,n,\; j=1,\cdots,n, \; i \not= j, \\
	J_{kk} := \sum_{i=1}^n e_i^{(k)} 
	\left( \int x_i^2 u(x,t)dx - \eta^{\rm PM}_{ii}(0) \right) 
	\equiv 0, \quad k=1,\cdots,n-1,
\end{eqnarray*}
where $e^{(k)}=(e_1^{(k)} \cdots e_n^{(k)}), k=1,\cdots,n-1$ are 
a set of $n-1$ basis vectors of the hyperplane 
${\cal H}=\{x \in {\bf R}^n | \sum_{i=1}^n x_i=0 \}$.
Particularly, a solution on the $N$-dimensional manifold ${\cal M}$ 
has $N-1$ constants of motion except the trivial one $J_0$.
This implies possibility that solutions on ${\cal M}$ may be 
explicitly solved by quadratures.
\end{remark}


Note that the m-projection $\hat u(x,t)$ of a solution $u(x,t)$ 
satisfies the PME (\ref{PME}) only when 
$\hat u(x,t)=u(x,t)$, in other words, $u(x,t)$ is a solution on the 
invariant manifold ${\cal M}$.
This is because the evolutional velocity of each m-projection $\hat u(x,t)$ 
along a common m-geodesic curve depends on 
how far from ${\cal M}$ the corresponding original solution $u(x,t)$ evolves. 
This phenomena is specific to the slow diffusion, and 
is quantitatively evaluated 
in terms of the expectation coordinates and the divergence as follows:


Let $\hat f_0(x) \in {\cal M}$ be the m-projection of the density $f_0(x)$.
Consider two solutions $u_1(x,t)$ and $u_2(x,t)$ of the PME satisfying 
$u_1(x,t_0)=f_0(x)$ and $u_2(x,t_0)=\hat f_0(x)$ at a certain time $t=t_0$.
Note that $u_2(x,t) \in {\cal M}$ for all $t$ because ${\cal M}$ is invariant.
From the moment conservation property of the m-projection 
stated in Proposition \ref{prop_proj}, 
the second moment matrices $H^{\rm PM}_i(t)$ of $u_i(x,t)$ for $i=1,2$
satisfy $H^{\rm PM}_1(t_0)=H^{\rm PM}_2(t_0)$.
However, their velocities at $t_0$ have the following relation:
\begin{eqnarray*}
	\dot H^{\rm PM}_1(t_0)-\dot H^{\rm PM}_2(t_0)
	&=& 2  \int f_0^m(x)-\hat f_0^m(x) dx \; I \\
	&=& 2m(m-1) 
	\left( {\cal I}[\hat f_0(x)]-{\cal I}[f_0(x)] \right)I
\end{eqnarray*}
by (\ref{sol_mom}) and the expression of the generalized entropy 
(\ref{dualent}). 
Using the relation in Remark \ref{rem_proj}, we have the following:
\begin{corollary}
\label{dependency}
Let $\hat f_0(x) \in {\cal M}$ be the m-projection of $f_0(x)$ and assume 
that two solutions $u_1(x,t)$ and $u_2(x,t)$ of the PME satisfy the conditions 
$u_1(x,t_0)=f_0(x)$ and $u_2(x,t_0)=\hat f_0(x)$ at a certain time $t=t_0$.
Then velocities of their respective second moment matrices at $t_0$ are
related by
\[
	\dot H^{\rm PM}_1(t_0)-\dot H^{\rm PM}_2(t_0)
	=2m(m-1){\cal D}[f_0(x) \| \hat f_0(x)]I.
\]
\end{corollary}

\begin{figure}[bhtp]
\begin{center}
  \includegraphics[width=12cm]{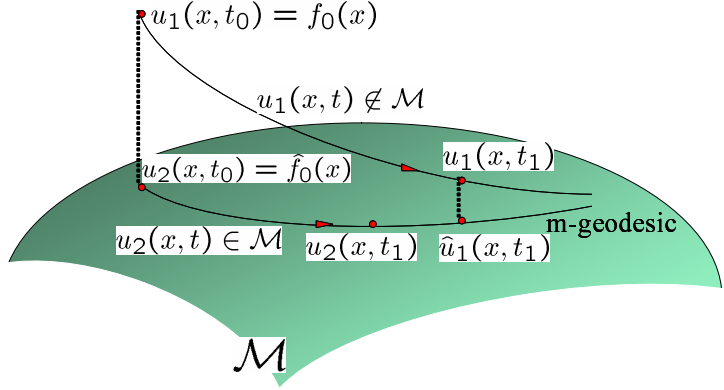}
\caption{For two solutions $u_1(x,t) \not\in {\cal M}$ and 
$u_2(x,t) \in {\cal M}$, the m-projection $\hat u_1(x,t)$ 
and $u_2(x,t)$ evolve along 
a common m-geodesic curve on ${\cal M}$ with different velocities.}
\label{Vsigmafig}
\end{center}
\end{figure}

Thus, the m-projection $\hat u_1(x,t)$ of $u_1(x,t) \not\in {\cal M}$, 
which has the common second moment matrix $H_1^{\rm PM}(t)$ for all $t$, 
evolves faster than $u_2(x,t) \in {\cal M}$,
while $\hat u_1(x,t)$ and $u_2(x,t)$ pass along a common m-geodesic curve 
on ${\cal M}$ by Theorem \ref{thm1}.
(See Figure \ref{Vsigmafig} and \cite{Ohara09} for numerical experiments.)
Finally the corollary suggests that by measuring the diagonal elements of 
$H_1^{\rm PM}(t)$ we can estimate how far
$u_1(x,t)$ is from ${\cal M}$ in terms of the divergence.
Note that the difference of velocities vanishes when $m \rightarrow 1$.
Hence, this is the specific property of the slow diffusions governed by 
the PME.

%

Next we study the behavior of the solutions $p(x,\tau)$ of 
the NFPE (\ref{DNFPE}).
Recall the transformation from a solution $u(x,t)$ of the PME to 
$p(x,\tau)$ given in Proposition \ref{prop_transfer}.
Since it holds that $dz=(t+1)^{-\alpha}\det (R) dx$, 
we have the relations of the moments:
\begin{eqnarray}
	\int p(z,\tau) dz &=& \int (t+1)^{\alpha}u(x,t)dz= 
	\det (R) \int u(x,t) dx \\
	\int z p(z,\tau) dz
		&=& (t+1)^{-\beta} \det (R) R \int x u(x,t) dx 
	\label{1stmom} \\
	\int zz^T p(z,\tau) dz &=& (t+1)^{-2\beta}\det (R) R 
		\left( \int xx^T u(x,t) dx \right) R^T
	\label{2ndmom}
\end{eqnarray}
The first relation shows that the solution of the NFPE also conserves 
its mass.
Let $\eta^{\rm FP}(\tau)$ and $H^{\rm FP}(\tau)$ be, respectively, 
the first and second moments of $p(x,\tau)$, i.e.,
\[
	\eta^{\rm FP}= (\eta_i^{\rm FP}), 
	\quad H^{\rm FP}=(\eta_{ij}^{\rm FP}),
\]
where
\[
	\eta_i^{\rm FP}:= \mathbf{E}_p[x_i],  
\quad   \eta_{ij}^{\rm FP}:=\mathbf{E}_p[x_ix_j]. 
\]
From the behavior of the moments of the PME and 
the above relations, we have 
\begin{eqnarray*}
	\eta^{\rm FP}(\tau)&=&(t+1)^{-\beta}\det (R)R \eta^{\rm PM}(t) 
	= e^{-\beta \tau}\det (R)R \eta^{\rm PM}(0) \\
	&=& e^{-\beta \tau}\eta^{\rm FP}(0), \\
	H^{\rm FP}(\tau)&=&(t+1)^{-2\beta}\det (R)RH^{\rm PM}(t)R^T \\
	&=&e^{-2\beta \tau}\det (R)[RH^{\rm PM}(0)R^T+\sigma^{\rm PM}_u(e^\tau-1)D] \\
	&=& e^{-2\beta \tau}H^{\rm FP}(0)
		+ e^{-2\beta \tau} \sigma^{\rm FP}_p(e^\tau-1)D,
\end{eqnarray*}
where the scaling $\tau=\ln(t+1)$ is assumed 
and $\sigma^{\rm FP}_p(t)$ is defined by
\begin{eqnarray*}
	\sigma^{\rm FP}_p(t)
		&:=& \det(R) \sigma^{\rm PM}_u(t) \\
		&=&2\int_0^{\ln(1+t)} d\tau' e^{\tau'+\alpha(1-m)\tau'} 
			\int p(x,\tau')^m dx 
\end{eqnarray*}
for a solution $u$ of the PME and the corresponding solution $p$ of the NFPE.
Note that differentiating the above by $t$, we have
\begin{equation}
	(1+t)^{\alpha(1-m)} \int p(x,\tau)^m dx
	= \det (R) \int u(x,t)^m dx.
\label{m_trans}
\end{equation}

For the limiting case $m \rightarrow 1$ (and accordingly 
$\beta \rightarrow 1/2$),
we see that the above expressions recover the well-known 
linear Fokker-Plank case with a drift vector $x/2$:
\[
		\eta^{\rm FP}(\tau)=e^{- \tau/2}\eta^{\rm FP}(0), \quad
		H^{\rm FP}(\tau)=e^{- \tau}H^{\rm FP}(0)
		+ 2(1-e^{-\tau})D.
\]
Since we know that $p(x,\tau)$ converges to $p_\infty(x) \in {\cal M}$ 
in (\ref{eq_density}) and it holds that
\begin{equation}
	\lim_{\tau \rightarrow \infty} H^{\rm FP}(\tau)
	=\sqrt{\det D} \left(\lim_{t \rightarrow \infty} 
	(t+1)^{-2\beta}\sigma^{\rm PM}_u(t) \right)D
\label{limitingcase}
\end{equation}
because $\det R=\sqrt{\det D}$, we conclude that 
the left-hand side of (\ref{limitingcase}) exists and 
$\sigma^{\rm PM}_u(t)=O(t^{2\beta})$ as $t \rightarrow \infty$ 
(Cf. Remark \ref{rem_thm}).
Summing up the above with Proposition \ref{prop_geod}, 
we obtain the following geometric property of the NFPE:
\begin{corollary}
Consider solutions of the NFPE 
with the common initial first and second moments.
Then their m-projections to ${\cal M}$ evolve along 
with the common m-geodesic curve from the density determined by 
the initial moments to the equilibrium $p_\infty(x)$.
\end{corollary}
Note that the following relation holds with the scaling $\tau=\ln(t+1)$:
\begin{equation}
	\frac{d}{d\tau}H^{\rm FP}(\tau)
	=(t+1)^{-2\beta} \left(-2\beta H^{\rm FP}(0)
		-2\beta \sigma_u^{\rm PM}(t)D+(t+1)
		 \frac{d \sigma_u^{\rm PM}(t)}{dt}D \right).
\label{viv}
\end{equation}
Hence, we cannot guarantee 
the monotonic behavior of the second moment matrix $H^{\rm FP}(\tau)$ 
unlike the linear Fokker-Planck equation.
For example, if the initial density $p(x,0)$ is not on 
${\cal M}$ but has the common second moments with the 
equilibrium density, we cannot expect the right-hand side of 
(\ref{viv}) is zero and the second moment matrix possibly oscillates 
around its equilibrium.

\subsection{Convergence rate of the solution of the PME to ${\cal M}$}
Finally, we show that the triangle equality of the divergence is useful to 
estimate the convergence rate of the solution of the PME to ${\cal M}$.
It is known \cite{CT,Otto,Toscani} that a solution of the NFPE decays exponentially 
with respect to the divergence, i.e.,
\begin{equation}
	{\cal D}[p(x,\tau) \| p_\infty(x)]
	={\cal F}[p(x,\tau)]-{\cal F}[p_\infty(x)]
	\le {\cal D}[p(x,0) \| p_\infty(x)]e^{-2 \beta \tau}.
\label{expdecay}
\end{equation}

\begin{proposition}
Let $u(x,t)$ be a solution of the PME and $\hat u(x,t)$ be the m-projection 
of $u(x,t)$ to the $q$-Gaussian family ${\cal M}$ at each $t$.
Then $u(x,t)$ asymptotically approaches ${\cal M}$ satisfying
\[
	{\cal D}[u(x,t) \| \hat u(x,t)] \le \frac{C_0}{1+t},
\]
where $C_0$ is a constant depending on the initial function $u(x,0)$.
\end{proposition}
Proof) Owing to the triangular equality of the m-projection in 
Proposition \ref{prop_proj}, it holds that
\[
	{\cal D}[p(x,\tau) \| \hat p(x,\tau)]
	+{\cal D}[\hat p(x,\tau) \| p_\infty(x)]
	={\cal D}[p(x,\tau) \| p_\infty(x)].
\]
Together with (\ref{expdecay}), we have
\[
	{\cal D}[p(x,\tau) \| \hat p(x,\tau)] 
		\le {\cal D}[p(x,0) \| p_\infty(x)]e^{-2 \beta \tau}.
\]

Let $u(x,t)$ and $\hat u(x,t)$ be functions defined 
from $p(x,\tau)$ and $\hat p(x,\tau)$, respectively, 
through the transformation in Proposition \ref{prop_transfer}.
Then, it is easy to see that
$\hat u(x,t) \in {\cal M}$ if and only if $\hat p(x,\tau) \in {\cal M}$ 
at each fixed $t$ (and $\tau$).
Further, since the first and second moments 
of $p(x,\tau)$ and $\hat p(x,\tau)$ 
are equal, so are those of $u(x,t)$ and $\hat u(x,t)$
from (\ref{1stmom}) and (\ref{2ndmom}).
Thus, we conclude that $\hat u(x,t)$ is also the m-projection of $u(x,t)$.  
It holds from (\ref{m_trans}) that
\[
	\det(R) \int \hat u(x,t)^m - u(x,t)^m dx
	= (1+t)^{\alpha(1-m)} \int \hat p(x,\tau)^m - p(x,\tau)^m dx. 
\]
Hence, the relation in Remark \ref{rem_proj} shows that 
\begin{eqnarray*}
	{\cal D}[u(x,t) \| \hat u(x,t)] 
	&=& (1+t)^{\alpha(1-m)}{\cal D}[p(x,\tau) \| \hat p(x,\tau)]/\det(R) \\
	& \le & (1+t)^{\alpha(1-m)-2\beta} C_0=(1+t)^{-1}C_0
\end{eqnarray*}
because $\alpha(1-m)-2\beta=-1$.
\hfill Q.E.D.

\begin{remark}
From the above result
we can conclude that the $L^1$ convergence 
of $u(x,t)$ to ${\cal M}$ is with the rate $1/\sqrt{1+t}$, using 
the Csiszar-Kullback inequality \cite{CT}.
This implies that the convergence to ${\cal M}$ 
is faster than that to the self-similar solution of the PME 
(cf. the next subsection), 
the rate of which is known to be $1/t^\beta$ when $1<m \le 2$ 
\cite{Vaz,Toscani}.
This might be a new aspect of the convergence properties for solutions of 
the PME.
\end{remark}

\subsection{The trajectory of self-similar solution}
\label{SSsol}
For the PME (\ref{PME}) there exists a special solution on ${\cal M}$ called 
the {\em self-similar solution} or {\em Barenblatt-Pattle solution} 
\cite{Bar,Pa}. 
The solution is expressed in terms of the $q$-Gaussian density 
for the case of unit mass by
\begin{eqnarray}
	u^{\rm BP}(x,t)&:=&
	t^{-\alpha} \left( C-\frac{(m-1)\beta}{2m}t^{-2\beta} x^Tx 
	\right)^{1/(m-1)}
	\nonumber \\
	&=&t^{-\alpha} 
	\exp_q \left(t^{-2\beta}x^T \Theta(1) x-\psi (0, \Theta(1)) \right) 
	\nonumber \\
	&=& \exp_q \left(x^T \Theta(t) x
		-\psi (0, \Theta(t)) \right)
	\nonumber \\
	&=& G_q(x;0,\Theta(t)), 
	\quad \Theta(t)=-t^{-1} \frac{\beta }{2m}I,
\label{self_sim}
\end{eqnarray}
where $C$ is a constant for $u^{\rm BP}(x,t)$ to have unit mass and 
the normalizing constant satisfies
\[
	\psi (0, \Theta(t))=\frac{1-Ct^{\alpha(1-m)}}{m-1}.
\]
The self-similar solution plays an important role in analysis of the PME 
\cite{Vaz}.
It is known that any solution for the PME with unit initial mass 
converges to $u^{\rm BP}(x,t)$, e.g., in $L^1$ norm 
$\lim_{t \rightarrow \infty} \|u(x,t)-u^{\rm BP}(x,t)\|_1=0$ \cite{Vaz,Toscani}.
Geometrically, it is also special in the following sense:
\begin{lemma}
\label{BPonM}
The trajectory of the self-similar solution 
$u^{\rm BP}(x,t)$ is a curve on ${\cal M}$ 
that is simultaneously an m- and e-geodesic.
\end{lemma}
Proof) 
In Theorem \ref{thm1} we have already proved that the trajectory is 
an m-geodesic.
Since (\ref{self_sim}) shows that the trajectory is expressed as 
a straight line in the canonical coordinate system $(\theta, \Theta)$, 
the statement follows from Proposition \ref{prop_geod}.
\hfill Q.E.D.
\begin{remark}
The property of the self-similar solution stated in the above lemma is called 
{\em doubly autoparallel} \cite{UO}.  
From this we can readily conclude that the trajectory of the self-similar 
solution is also a geodesic with respect to the Levi-Civita connection.
\end{remark}

As an application of this property, we have the following decomposition 
of the divergence:
\begin{proposition}
Let $u(x,t)$ be a solution of the PME and $\hat u(x,t)$ be its m-projection to 
${\cal M}$.
For the trajectory of the self-similar solution $u^{\rm BP}(x,t)$ with $0<t$ 
denoted by $\gamma$, define the m-projection of $\hat u(x,t)$ to $\gamma$ by
\[
	u^*(x,t):=\arg \min_{r(x) \in \gamma} {\cal D}[\hat u(x,t) \| r(x)].
\]
Then it holds for all $t >0$ that 
\begin{eqnarray*}
	{\cal D}[u(x,t) \| u^{\rm BP}(x,t)]
	&=&{\cal D}[u(x,t) \| \hat u(x,t)]
	\nonumber \\
	&&+{\cal D}[\hat u(x,t) \| u^*(x,t)]
	+{\cal D}[u^*(x,t) \| u^{\rm BP}(x,t)].
\end{eqnarray*}
\end{proposition}
Proof) From Proposition \ref{prop_proj}, we have 
\[
	{\cal D}[u(x,t) \| u^{\rm BP}(x,t)]
	={\cal D}[u(x,t) \| \hat u(x,t)]
		+{\cal D}[\hat u(x,t) \| u^{\rm BP}(x,t)].
\]
The decomposition 
\[
	{\cal D}[\hat u(x,t) \| u^{\rm BP}(x,t)]
		={\cal D}[\hat u(x,t) \| u^{\rm BP}(x,t)]
		+{\cal D}[u^*(x,t) \| u^{\rm BP}(x,t)]
\]
follows from the standard argument of the Pythagorean relation 
in information geometry \cite{Amari,AN}.
\hfill Q.E.D.

\section{Conclusions}

We have studied the behavior of the solutions to the PME and NFPE 
focusing on the $q$-Gaussian family ${\cal M}$.
By proving that ${\cal M}$ is a stable invariant manifold of the both 
equations, 
we have obtained several properties of the solutions, e.g,
$N(={\rm dim}{\cal M})$ constants of motions, 
the convergence rate to ${\cal M}$ and geometrical characterization 
of the self-similar solution of the PME.
In particular,
the dependency of the evolutional velocity on the divergence 
from ${\cal M}$ (Corollary \ref{dependency}) 
would be a peculiar phenomena to the slow diffusion.

Through the analysis, we see that the 
generalized concepts of statistical physics and the compatibly defined 
information geometric structure on ${\cal M}$ 
provide us with abundant and precise information on 
the behavior of solutions.

In \cite{Otto}, Otto reported that the PME can be regarded as a gradient 
system via Riemannian geometry with the Wasserstein metric.
The relation with the framework in the present paper is left unclear.
Another important future work would be to confirm how 
the obtained results are analogously extended to the other parameter 
ranges: $2 \le m$ or $m<1$ (fast diffusion), 
or the other type of nonlinear diffusion equation.

\ack
The authors wish to thank Prof. A. Fujiwara, 
Prof. T. Kurose and Prof. M. Tanaka for their helpful comments.


\section*{References}
\begin{thebibliography}{9}
 \bibitem{Mu}	Muskat M 1937
		{\it The Flow of Homogeneous Fluids Through Porous Media}
		(New York: McGraw-Hill)
 \bibitem{Buck} Buckmaster J 1977 
		Viscous sheets advancing over dry beds
		{\it J. Fluid Mech.} {\bf 81} 735--56
 \bibitem{LP}	Larsen E W and Pomraning G C 1980
		Asymptotic analysis of nonlinear Marshak waves
		{\it SIAM J. Appl. Math.} {\bf 39} 201--12
 \bibitem{Kath} Kath W L 1985
		Waiting and propagating fronts in nonlinear diffusion
		{\it Physica D} {\bf 12} 375--81
 \bibitem{Bar2} Barenblatt G I 1996 
		{\it Scaling, self-similarity and intermediate asymptotics} 
		(Cambridge: Cambridge Univ. Press)
 \bibitem{FK}	Friedman A and Kamin S 1980
		The asymptotic behavior of gas in an N-dimensional 
		porous medium
		{\it Trans. Amer. Math. Soc.} {\bf 262} 551--63
 \bibitem{Ar}	Aronson D G 1985 
		{\it The Porous Medium Equation} 
		Lecture Notes in Mathematics 1224 
		(Berlin/New York: Springer-Verlag)
 \bibitem{Newman} Newman W 1984
		A Lyapunov functional for the evolution of solutions to 
		the porous medium equation to self-similarity I
		{\it J. of Math. Phys.} {\bf 25} 3124--27
 \bibitem{Ralston} Ralston J 1984
		A Lyapunov functional for the evolution of solutions to 
		the porous medium equation to self-similarity II
		{\it J. of Math. Phys.} {\bf 25} 3120--23
 \bibitem{PlaPla} Plastino A R and Plastino A 1995 
		Non-extensive statistical mechanics and generalized 
		Fokker-Planck equation
		{\it Physica A} {\bf 222} 347--54
 \bibitem{Shiino} Shiino M 1998
		H-theorem with generalized relative entropies and 
		the Tsallis statistics
		{\it J. Phys. Soc. Jpn.} {\bf 67} 3658--60
 \bibitem{Frank01} Frank T D 2001
		Lyapunov and free energy functionals of generalized
		Fokker-Planck equations
		{\it Physics Letters A} {\bf 290} 93--100
 \bibitem{Frank02} Frank T D 2002
		Generalized Fokker-Planck equations derived from 
		generalized linear nonequilibrium thermodynamics
		{\it Physica A} {\bf 310} 397--412
 \bibitem{CT}	Carrillo J A and Toscani G 2000 
		Asymptotic L1-decay of solutions of the porous medium 
		equation to self-similarity
		{\it Indiana Univ. Math. J.} {\bf 49} 113--41
 \bibitem{Otto}	Otto F 2001
		The geometry of dissipative evolution equations: 
		the porous medium equation
		{\it Comm. Partial Differential Equations} {\bf 26} 101--74
 \bibitem{Vaz}	V\'azquez J L 2003
		Asymptotic behaviour for the porous medium equation posed 
		in the whole space
		{\it J. Evol. Equ.} {\bf 3} 67--118
 \bibitem{Toscani} Toscani G 2005
		A central limit theorem for solutions of the porous medium 
		equation
		{\it J. Evol. Equ.} {\bf 5} 185--203
 \bibitem{Frank05} Frank T D 2005
		{\it Nonlinear Fokker-Planck Equations: 
		Fundamentals and Applications} 
		(Berlin/Heidelberg: Springer-Verlag) 
 \bibitem{Vaz06}V\'azquez J L 2006
		{\it Smoothing and decay estimates for nonlinear diffusion 
		equations : equations of porous medium type}
		(Oxford: Oxford University)
 \bibitem{Vaz07} V\'azquez J L 2007
		{\it The porous medium equation: Mathematical theory}
		(Oxford: Oxford University)
 \bibitem{AGS} Ambrosio L A, Gigli N and Savar\'e G 2005
		{\it Gradient flows: in metric spaces and 
		in the space of probability measures}
		(Basel: Birkh\"auser)
 \bibitem{CFT} Carrillo J A, Di Francesco M and Toscani G 2006
		Intermediate asymptotics beyond homogeneity 
		and self-similarity: long time behavior for $u_1=\Delta \phi(u)$		{\it Arch. Ration. Mech. Anal.} {\bf 180} 127--49
 \bibitem{DP} Dolbeault J and del Pino M 2002
		Best constants for gagliardo-nirenberg inequalities and 
		application to nonlinear diffusions 
		{\it J. Math. Pures Appl.} {\bf 81} 847--75
 \bibitem{Villani} Villani C 2003
		{\it Topics in optimal transportation}
		(Providence: American Mathematical Society)
 \bibitem{CMV} Carrillo J A, McCann R J and Villani C 2006
		Contractions in the 2-Wasserstein Length Space and 
		Thermalization of Granular Media
		{\it Arch. Ration. Mech. Anal.} {\bf 179} 217--63
 \bibitem{OW} Otto F and Westdickenberg M 2006
		Eulerian calculus for the contraction in thewasserstein
		distance
		{\it SIAM J. Math. Anal.} {\bf 37} 1227--55
 \bibitem{Villani2} Villani C 2009
		{\it Optimal transport: old and new}
		(Berlin/Heidelberg: Springer-Verlag)
 \bibitem{TLSM} Tsallis C, Levy S V F, Souza A M C and Maynard R 1995
		Statistical-mechanical foundation of the ubiquity of Levy 
		distributions in nature
		{\it Physical Review Letters} {\bf 75} 3589--93, 
		1996 Errata, {\bf 77} 5442
 \bibitem{ST}	Suyari H and Tsukada M 2002
		Law of errors in Tsallis statistics
		{\it IEEE Trans. Inf. Theory} {\bf 51} 2, 753--7
 \bibitem{Amari} Amari S-I 1985
		{\it Differential-Geometrical Methods in Statistics}
		{\it Lecture Notes in Statistics}, Vol.~28, 
		(New York: Springer-Verlag)
 \bibitem{AN} 	Amari S-I and Nagaoka H 2000
		{\it Methods of Information Geometry}
		Trans. Math. Mono., Vol.191, (Providence: AMS)
 \bibitem{Eguchi} Eguchi S 2006,
		Information geometry and statistical pattern recognition
		{\it Sugaku Expositions} {\bf 19} 197--216,
		(originally 2004 {\it S\=ugaku}, {\bf 56} 380--99 
		{\it in Japanese}.)
 \bibitem{MTKE}	Murata N, Takenouchi T, Kanamori T and Eguchi S 2004
		Information geometry of U-Boost and Bregman divergence
		{\it Neural Computation} {\bf 16} 1437--81
 \bibitem{Naudts02} Naudts J 2002
		Deformed exponentials and logarithms in generalized
		thermodynamics
		{\it Physica A} {\bf 316} 323--34
 \bibitem{Naudts041} Naudts J 2004
		Generalized thermodynamics based on deformed exponentials and
		logarithmic functions
		{\it Physica A} {\bf 340} 32-40
 \bibitem{Naudts042} Naudts J 2004
		Continuity of a class of entropies and relative entropies,
		{\it Reviews in Mathematical Physics} {\bf 16} 809--22
 \bibitem{Naudts043} Naudts J 2004
		Estimators, escort probabilities, and $\phi$-exponential
		families in statistical physics
		{\it J. Ineq. Pure Appl. Math} {\bf 5} 102
 \bibitem{Burb} Burbea J 1986
		Informative geometry of probability spaces
		{\it Expo. Math.} {\bf 4} 347--78
 \bibitem{Mit}  Mitchell A F S 1989
		The information matrix, skewness tensor and 
		$\alpha$-connections for the general multivariate 
		elliptic distribution
		{\it Ann. Inst. Statist. Math.} {\bf 41} 289--304
 \bibitem{OC} Oller J M and Corcuera J M 1995 
		Intrinsic analysis of statistical estimation 
		{\it Ann. Stat.} {\bf 23} 1562--81
 \bibitem{Andai} Andai A 2008
		On the geometry of generalized Gaussian distributions 
		{\it J. Multivariate Anal.} {\bf 100} 777-93
 \bibitem{GD} 	Grunwald P D and David A P 2004
 		Game theory, maximum entropy, minimum discrepancy and 
		robust baysian decision theory
		{\it Annals of Statistics} {\bf 32} 1367--433
 \bibitem{David} David A P 2007
 		The geometry of proper scoring rules,
		{\it Annals of Institute of Statistics} {\bf 59}, 77-93
 \bibitem{ME} 	Minami M and Eguchi S 2002
		Robust blind source separation by beta-divergence
		{\it Neural Computation} {\bf 14}, 1859-86
 \bibitem{HE}	Higuchi I and Eguchi S 2004
		Robust principal component analysis with adaptive selection 
		for Tuning Parameters,
		{\it J. Machine Learning Research} {\bf 5}, 453--71
 \bibitem{TE} 	Takenouchi T and Eguchi S 2004
 		Robustifying AdaBoost by adding the naive error rate
 		{\it Neural Computation} {\bf 16} 767--87
 \bibitem{Tsallis1} Tsallis C 1988
		Possible generalization of Boltzmann-Gibbs statistics,
		{\it J. Stat. Phys.} {\bf 52} 479--87
 \bibitem{WS05} Wada T and A.M.Scarfone A M 2005
		Connection between Tsallis' formalisms employing the standard 			linear average energy and ones employing the normalized 
		$q$-average energy
		{\it Physics Letters A} {\bf 335} 351--62
 \bibitem{Callen} Callen H B 1985
		{\it Thermodynamics and an Introduction to Thermostatistics
		second ed.}
		(New York: Wiley)
 \bibitem{Ohara09} Ohara A 2009
		Geometric study for the Legendre duality of generalized 
		entropies and its application to the porous medium equation, 
		{\it Topical issue on SigmaPhi 2008,  
		European Physical Journal B} {\bf 70} 15-28
 \bibitem{Bar}	Barenblatt G I 1952
		On some unsteady motions of a liquid or a gas 
		in a porous medium
		{\it Prikl. Mat. Mekh.} {\bf 16} 67-78 (in Russian)
 \bibitem{Pa}	Pattle R E 1959
		Diffusion from an instantaneous point source with 
		concentration dependent coefficient
		{\it Quart. Jour. Mech. Appl. Math.} {\bf 12} 407--09
 \bibitem{UO} 	Uohashi K and Ohara A 2004
		Jordan Algebras and Dual Affine Connections on
		Symmetric Cones
		{\it Positivity} {\bf 8} 369--78
\end{thebibliography}
\end{document}